\documentclass[floats,aps,showpacs,twocolumn,superscriptaddress]{revtex4-1}

\usepackage[dvips]{graphicx}
\usepackage{amsfonts}
\usepackage{amsmath,amssymb}
\usepackage{dsfont}
\usepackage{placeins}
\usepackage{afterpage}
\usepackage{flafter}
\usepackage[usenames,dvipsnames]{color}

\renewcommand{\Im}{\operatorname{Im}}
\renewcommand{\Re}{\operatorname{Re}}
\newcommand{\Ps}[1]{P_{\mathrm{#1}}}                    
\newcommand{\ps}[2]{p_{\mathrm{#2}}(#1)}

\newcommand{\ket}[1]{\left| #1 \right\rangle}
\newcommand{\bra}[1]{\left\langle #1 \right|}
\newcommand{\scal}[2]{\langle #1 | #2 \rangle}

\newcommand{\op}[1]{\hat{#1}}
\newcommand{\mat}[1]{#1}

\newcommand{\tauch}{\tau_{\mathrm{H,c}}}
\newcommand{\heff}{h_{\mathrm{eff}}}
\newcommand{\heffbar}{\hbar_{\mathrm{eff}}}
\newcommand{\Areg}{A_{\mathrm{r}}}
\newcommand{\Nreg}{N_{\mathrm{r}}^{\mathrm{sc}}}
\newcommand{\Ntil}{N^{\mathrm{f}}_{\mathrm{r}}}
\newcommand{\Nch}{N_{\mathrm{c}}^{\mathrm{sc}}}
\newcommand{\rhorsc}{\rho_{\mathrm{r}}^{\mathrm{sc}}}
\newcommand{\rhochsc}{\rho_{\mathrm{c}}^{\mathrm{sc}}}
\newcommand{\rhoch}{\rho^{\mathrm{f}}_{\mathrm{c}}}
\newcommand{\rhor}{\rho^{\mathrm{f}}_{\mathrm{r}}}

\newcommand{\dch}{\Delta_{\mathrm{c}}}
\newcommand{\mmax}{m_{\mathrm{max}}}

\newcommand{\im}{\operatorname{i}}
\newcommand{\e}{\operatorname{e}}
\newcommand{\dd}{\mathrm{d}}

\newcommand{\hf}{{\mathrm{sym}}}

\begin{document}

\title{Consequences of Flooding on Spectral Statistics}

\author{Torsten Rudolf}
\affiliation{Institut f\"ur Theoretische Physik, Technische Universit\"at
             Dresden, 01062 Dresden, Germany}
\affiliation{Max-Planck-Institut f\"ur Physik komplexer Systeme, N\"othnitzer
Stra\ss{}e 38, 01187 Dresden, Germany}

\author{Normann Mertig}
\affiliation{Institut f\"ur Theoretische Physik, Technische Universit\"at
             Dresden, 01062 Dresden, Germany}
\affiliation{Max-Planck-Institut f\"ur Physik komplexer Systeme, N\"othnitzer
Stra\ss{}e 38, 01187 Dresden, Germany}

\author{Steffen L\"ock}
\affiliation{Institut f\"ur Theoretische Physik, Technische Universit\"at
             Dresden, 01062 Dresden, Germany}

\author{Arnd B\"acker}\
\affiliation{Institut f\"ur Theoretische Physik, Technische Universit\"at
             Dresden, 01062 Dresden, Germany}
\affiliation{Max-Planck-Institut f\"ur Physik komplexer Systeme, N\"othnitzer
Stra\ss{}e 38, 01187 Dresden, Germany}

\date{\today}

\begin{abstract}

We study spectral statistics in systems with a 
mixed phase space, in which regions of regular and chaotic motion coexist. 
Increasing their density of states, we observe a transition of the
level-spacing distribution $P(s)$ from Berry-Robnik to Wigner statistics,
although the underlying classical phase-space structure 
and the effective Planck constant $\heff$ remain unchanged. 
This transition is induced by flooding, i.e., the disappearance of 
regular states due to increasing regular-to-chaotic couplings.
We account for this effect by a flooding-improved Berry-Robnik 
distribution, in which an effectively reduced size of the regular island enters. 
To additionally describe power-law level repulsion at small spacings, we
extend this prediction by explicitly considering the tunneling 
couplings between regular and chaotic states. 
This results in a flooding- and tunneling-improved Berry-Robnik distribution
which is in excellent agreement with numerical data.

\end{abstract}

\maketitle
\noindent

\section{Introduction}

\begin{figure}[!b]
  \begin{center}
    \includegraphics[width=\linewidth]{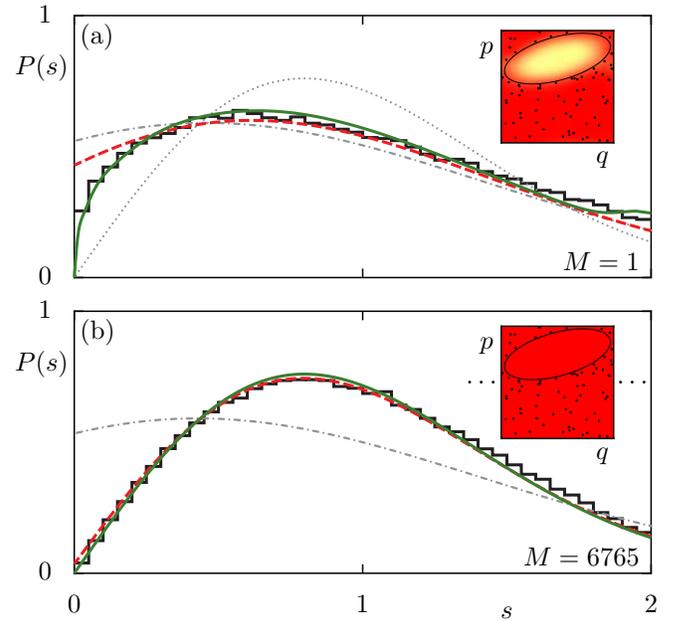}
    \caption{(Color online) Level-spacing distribution $P(s)$ of the model
      system (see Sec.~\ref{sec:modelSystem}) for $\heff\approx1/13$. The 
      numerical data (black histogram) is compared to the flooding-improved
      Berry-Robnik distribution (red dashed lines), 
      Eq.~(\ref{eq:effBerryRobnik}), as well as to the flooding- and 
      tunneling-improved Berry-Robnik distribution (green solid lines), 
      Eq.~(\ref{eq:ftibr_dist}), for system sizes (a) $M=1$ (weak flooding) 
      and (b) $M=6765$ (strong flooding); $M$ is introduced in 
      Sec.~\ref{sec:kickedSystems}. 
      For comparison the Wigner distribution (dotted lines) and the 
      Berry-Robnik distribution (dash-dotted lines) are shown. 
      The insets show averaged Husimi functions of chaotic eigenstates.}
    \label{fig:introPS}
  \end{center}
\end{figure}

The universal relation between the statistics of quantum spectra and classical
mechanics is a fundamental cornerstone of quantum chaos: 
For systems with regular dynamics it was conjectured that spectral statistics 
show Poissonian behavior \cite{BeTa1977}. 
In contrast, systems with chaotic dynamics should be described by random 
matrix theory \cite{BoGiSc1984, CaVaGu1980}, which can be explained in terms of
periodic orbits \cite{Be1985, SiRi2001, HeMuAlBrHa2007}.
For generic Hamiltonian systems with a mixed phase space, in which disjoint 
regions of regular and chaotic motion coexist, universal spacing statistics 
were obtained by Berry and Robnik \cite{BeRo1984}.
Their derivation is based on the semiclassical eigenfunction hypothesis 
\cite{Pe1973, Be1977, Vo1979}, which states that eigenfunctions of a quantum 
system semiclassically localize on those regions in phase space a typical that
orbit explores in the long-time limit.
For regular states in one-dimensional systems this corresponds to the WKB 
quantization condition \cite{LaLi1991, BeBaTaVo1979}
\begin{align}
  \label{eq:bohr-sommerfeld}
  \oint_{\mathcal{C}_{m}} p \, \dd{}q = \heff \left( m + \frac{1}{2} \right).
\end{align}
It shows that the regular state, labeled by the quantum number $m$, localizes 
on the quantizing torus $\mathcal{C}_{m}$ which encloses the area 
$\heff \left( m + \frac{1}{2} \right)$ in phase space.
On the other hand the semiclassical eigenfunction hypothesis implies that
chaotic states uniformly extend over the chaotic region of phase
space. Assuming that the disjoint regular and chaotic regions give rise to 
statistically uncorrelated level sequences, one obtains the Berry-Robnik 
level-spacing distribution \cite{BeRo1984}; see Fig.~\ref{fig:introPS} 
(dash-dotted lines).

The assumption of uncorrelated regular and chaotic level sequences does not 
hold in the presence of dynamical tunneling 
\cite{DaHe1981, HaOtAn1984, BoToUl1993, ToUl1994, ShIk1995, BrScUl2001, 
BrScUl2002, ShFiGuRe2006, BaeKetLoeSch2008, LoBaKeSc2010, BaKeLo2010, 
DeGrHeHoReRi2000, StOsWiRa2001, He2001}, 
which quantum mechanically couples regular and chaotic states. 
If such tunneling couplings are small, regular eigenstates will typically have
tiny chaotic admixtures and vice versa. The influence of such weak couplings on
spacing statistics can be described perturbatively 
\cite{Le1993, PoNa2007, ViStRoKuHoGr2007, BaRo2010, BaKeLoMe2011}. 
Based on this description a tunneling-improved Berry-Robnik 
distribution was derived recently, which explains the power-law distribution 
of small spacings in mixed systems \cite{BaKeLoMe2011}.

For systems with a large density of states, it is observed 
\cite{BoToUl1990,BoToUl1993,BaKeMo2005, BaKeMo2007, FeBaKeRoHuBu2006, 
IsTaSh2010} that a regular WKB state strongly couples to many chaotic states. 
As a consequence, the corresponding regular eigenstate disappears
and chaotic eigenstates penetrate into the regular island, ignoring 
the semiclassical eigenfunction hypothesis.
This effect is called flooding \cite{BaKeMo2005, BaKeMo2007}.
It causes the number $\Ntil$ of regular eigenstates that actually exist in 
the regular island to be smaller than the number $\Nreg$ expected from the 
semiclassical eigenfunction hypothesis.
In Refs.~\cite{BaKeMo2005, BaKeMo2007} it was found that in addition to the 
WKB quantization condition \eqref{eq:bohr-sommerfeld} the regular state on 
the $m$th quantizing torus exists only if
\begin{align}
  \label{eq:exCrit2}
  \gamma_{m} < \frac{1}{\tauch}.
\end{align}
Here, $\gamma_{m}$ is the tunneling rate, which describes the initial 
exponential decay of the $m$th WKB state to the chaotic region.
The Heisenberg time $\tauch=\heff/\dch$ is the ratio of the effective Planck 
constant $\heff$ and the mean level spacing of the chaotic spectrum $\dch$.

In this paper we study the consequences of flooding on spectral statistics in 
systems with a mixed phase space. 
With increasing density of states we observe a transition of the level-spacing 
distribution from Berry-Robnik [see Fig.~\ref{fig:introPS}(a)] to 
Wigner statistics [see Fig.~\ref{fig:introPS}(b)], although the underlying 
classical phase-space structure and $\heff$ remain unchanged.
This transition is demonstrated quantitatively for model systems with a simple
phase-space structure, but it is expected to hold for generic systems with a mixed phase
space. In order to explain the transition, we introduce a flooding-improved 
Berry-Robnik distribution which takes into account that only 
$\Ntil \le \Nreg$ regular states survive in the regular region.
We find good agreement with numerical data; see Fig.~\ref{fig:introPS} (red 
dashed lines). 
We unify this intuitive prediction with the tunneling-improved 
Berry-Robnik distribution \cite{BaKeLoMe2011}, which explicitly considers 
the tunneling couplings between regular and chaotic states.
This results in a tunneling- and flooding-improved Berry-Robnik distribution, 
which excellently reproduces the 
observed transition from Berry-Robnik to Wigner statistics as well as 
the power-law level repulsion at small spacings; 
see Fig.~\ref{fig:introPS} (green solid lines).

This paper is organized as follows: In Sec.~\ref{sec:modelSystem} we introduce 
a family of model systems. Their level-spacing distribution is studied in 
Sec.~\ref{sec:specStatWithFlooding}, where we demonstrate the transition from 
Berry-Robnik to Wigner statistics numerically and explain it by the flooding 
of regular states. We conclude with a summary in Sec.~\ref{sec:outlook}.

\section{Model System}
\label{sec:modelSystem}

In this section we introduce a family of model systems for which the 
consequences of flooding can be studied in detail. 

\subsection{Classical dynamics}
\label{sec:kickedSystems}

\begin{figure}[b]
  \begin{center}
    \includegraphics[width=\linewidth]{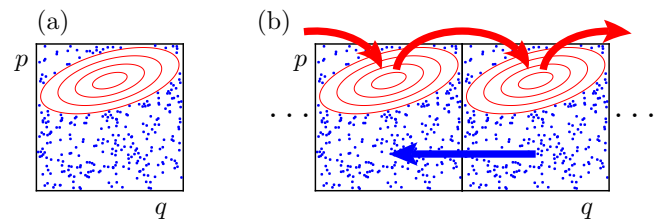}
    \caption{(Color online) Phase-space portrait of the model system, 
      Eq.~(\ref{eq:strobMap}). For one unit cell $M=1$ (a) the regular island 
      (red lines) is embedded in the chaotic sea (blue dots). 
      For systems with $M>1$ (b) the phase space consists of $M$ such unit 
      cells side by side. 
      The arrows indicate the transport in the regular islands and in the 
      chaotic sea.}
    \label{fig:phasespace}
  \end{center}
\end{figure}

We consider systems with a mixed phase space where classically disjoint 
regions of regular and chaotic motion coexist. As examples we choose 
one-dimensional kicked systems, described by the classical Hamilton function
\begin{align}
  \label{eq:kickedHamiltonian}
  H(q,p,t) = T(p) + V(q) \sum_{n \in \mathbb{Z}} \delta(t-n),
\end{align}
where $T(p)$ is the kinetic energy and the potential $V(q)$ is applied once 
per kicking period. The dynamics of such systems is determined by the 
stroboscopic mapping $\mathcal{M}$ of the positions and the momenta 
$(q_{n}, p_{n})$ at times $t = n$ just after each kick \cite{Ha2001},
\begin{align}
  \label{eq:strobMap}
  \mathcal{M}: (q_{n+1}, p_{n+1}) = (q_{n} + T'(p_{n}), p_{n} - V'(q_{n+1})).
\end{align}

We design the example systems similar to 
Refs.~\cite{HuKeOtSch2002,BaKeMo2005,BaeKetLoeSch2008,BaKeLo2010} 
by the piecewise linear functions
\begin{align}
  \label{eq:piecelinfunc_t}
  t'(p) &= \begin{cases} 
    -1 + s_{1} (p + 1/4)  & \mathrm{for }\ \ p \in \ ]-1/2, 0[,\\
    +1 - s_{2} (p - 1/4)  & \mathrm{for }\ \ p \in \ ]0, 1/2[,
  \end{cases}\\
  \label{eq:piecelinfunc_v}
  v'(q) &= -rq -(1-r) \lfloor q + 1/2 \rfloor,
\end{align}
where $\lfloor x \rfloor$ is the floor function and $t'(p)$ is periodically
extended. 
Smoothing the functions $t'(p)$ and $v'(q)$ with a Gaussian
$G_{\epsilon}(z) = \exp(-z^{2}/2\epsilon^{2})/\sqrt{2\pi\epsilon^{2}}$, one 
obtains analytic functions
\begin{align}
  \label{eq:anaFunc_T}
  T'(p) &= \int_{-\infty}^{\infty} \dd{}z\, t'(z) G_{\epsilon}(p-z),\\
  \label{eq:anaFunc_V}
  V'(q) &= \int_{-\infty}^{\infty} \dd{}z\, v'(z) G_{\epsilon}(q-z).
\end{align}
By construction these functions have the periodicity properties
\begin{align}
  \label{eq:periodPropT}
  T'(p+k) &= T'(p),\\
  \label{eq:periodPropV}
  V'(q+k) &= V'(q) - k,
\end{align}
for $k \in \mathbb{Z}$. 
This allows to consider the map $\mathcal{M}$ on
a torus, i.e., $(q, p) \in [-M/2, M/2[ \times [-1/2, 1/2[$ with periodic
boundary conditions and $M \in \mathbb{N}$. 
Due to the choice of $T'(p)$ and $V'(q)$, the dynamics is equivalent in each 
unit cell of phase space with $q \in [k-1/2, k+1/2[$ and $k \in \mathbb{Z}$; 
see Fig.~\ref{fig:phasespace}. 
In the following we choose the parameters 
$s_{1}\in[5,20]$, $s_{2}=2$, $r=0.46$, and $\epsilon=0.005$ such that each 
unit cell has a regular island centered at $(\bar{q}_{k}, \bar{p}) = (k, 1/4)$.
The area of one such island is $\Areg \approx 0.32$, which equals the 
relative size of the regular region in phase space.

Since the islands are transporting to the next unit cell in the positive
$q$ direction, i.e., $\mathcal{M}(\bar{q}_{k}, \bar{p}) = (\bar{q}_{k+1}, 
\bar{p})$, the center of each island is a fixed point of the $M$th iterate of 
the map, $\mathcal{M}^{M}(\bar{q}_{k}, \bar{p}) = (\bar{q}_{k}, \bar{p})$. 
The surrounding chaotic sea has an average drift in the negative $q$ direction 
as the overall transport of the system is zero 
\cite{DiKeOtSch2001, DiKeSch2005}; see Fig.~\ref{fig:phasespace}.
Quantum mechanically this transport suppresses the localization of 
chaotic eigenstates.
In our model systems the hierarchical regions
around the regular islands are sufficiently small, and also the
effects of partial transport barriers and nonlinear resonance chains 
are irrelevant to the numerical studies.

\subsection{Quantization}

The quantum system is given by the time-evolution operator over 
one period of the driving,
\begin{align}
  \label{eq:timeOp}
  \op{U} = \exp\left(-\frac{\im}{\heffbar}V(\op{q})\right) 
              \exp\left(-\frac{\im}{\heffbar}T(\op{p})\right);
\end{align}
see, e.g., Refs.~\cite{BeBaTaVo1979, BeHa1980, ChSh1986}. Quantizing the map
$\mathcal{M}$ on a two-torus induces the Bloch phases $\theta_{q}$ and 
$\theta_{p}$ \cite{KeMeRo1999, ChSh1986} which characterize the 
quasi-periodicity conditions on the torus. 
The Bloch phase $\theta_{p}$ is limited by 
$M\left( \theta_{p} + N/2 \right) \in \mathbb{Z}$ because of the periodic 
boundary conditions, whereas $\theta_{q} \in [0, 1[$ can be chosen arbitrarily
\cite{KeMeRo1999, BaKeMo2007}.

Due to the quantization on a compact torus the effective Planck constant 
$\heff = 2 \pi \heffbar$ is determined by the number of unit cells $M$ and the 
dimension of the Hilbert space $N$,
\begin{align}
  \label{eq:heff}
  \heff = \frac{M}{N}.
\end{align}
Here $N \in \mathbb{N}$ is a free parameter of the quantization and the 
semiclassical limit is reached for $\heff \to 0$. 
Note that $M$ and $N$ are chosen by continued fractions of 
$\heff = 1/(d + \sigma)$ with $\sigma = (\sqrt{5}-1)/2$ being the golden mean 
and $d \in \mathbb{N}$. 
This ensures that $\heff = M/N$ is as irrational as possible \cite{BaKeMo2005}. 
If $M$ and $N$ were commensurate the quantum system would effectively be reduced
to less than $M$ cells. In the following we choose $d=12$, leading 
to $(M, N) = (1, 13), (21, 265), (610, 7697), (6765, 85361)$, 
such that the effective Planck constant is approximately fixed at 
$\heff \approx 1/13$.
 
The eigenvalue equation
\begin{align}
  \label{eq:eigvalProb}
  \op{U} \ket{\phi_{n}} = \e^{\im \phi_{n}} \ket{\phi_{n}}
\end{align}
gives $N$ eigenphases $\phi_{n} \in [0, 2\pi[$ with corresponding eigenvectors 
$\ket{\phi_{n}}$. 
For fixed $\heff$ it is possible to tune the density of 
states by varying $M$ and $N$, i.e., for increasing $M, N$ with approximately constant
$\heff=M/N$ the density of states rises and flooding becomes more and more prominent,
as will be discussed in Sec.~\ref{sec:flooding}.
In order to numerically solve the eigenvalue equation \eqref{eq:eigvalProb} for 
$N > 10^{4}$ we use a band-matrix algorithm, 
see the Appendix.

\section{Spectral Statistics And Flooding}
\label{sec:specStatWithFlooding}

In this section we study the consequences of flooding on spectral statistics. 
In Sec.~\ref{sec:br} we consider the model systems introduced in 
Sec.~\ref{sec:modelSystem}. Increasing their density of states ($M\to\infty$) 
at fixed $\heff$ gives the flooding limit for which
we obtain a transition of the 
level-spacing distribution $P(s)$ from Berry-Robnik to Wigner statistics.
In Sec.~\ref{sec:flooding} we discuss flooding of regular states.
Based on this discussion, we introduce the flooding-improved Berry-Robnik 
distribution $P_{\mathrm{fi}}(s)$ in 
Sec.~\ref{sec:effBeRoStat}, which intuitively explains how the flooding of 
regular states causes the transition from Berry-Robnik to Wigner statistics. 
In Sec.~\ref{sec:complTheory} we unify this prediction with the results of
Ref.~\cite{BaKeLoMe2011}, leading to the more sophisticated flooding- and 
tunneling-improved Berry-Robnik distribution $P_{\mathrm{fti}}(s)$. 
This distribution additionally accounts for the effects of level repulsion 
between regular and chaotic states.
In Sec.~\ref{sec:compare} we consider three limiting cases in which 
level repulsion vanishes. In particular we discuss that the 
semiclassical limit, $\heff\to 0$ with fixed $M$, leads to the
standard Berry-Robnik statistics, while Wigner statistics are obtained 
in the flooding limit considered in this paper.

\subsection{Spacing statistics of the model system}
\label{sec:br}

We investigate the spectral statistics of the model systems introduced in 
Sec.~\ref{sec:modelSystem} numerically. 
In order to increase the statistical significance of the spectral data, we 
perform ensemble averages by varying the parameter $s_{1}$ of the map; see 
Eq.~\eqref{eq:piecelinfunc_t}.
This modifies the chaotic dynamics but leaves the dynamics of the regular 
region unchanged. 
Also the Bloch phase $\theta_{q}$ is used for ensemble averaging.
For the parameters $(M, N) = (1, 13), (21, 265), 
(610, 7697), (6765, 85361)$ we choose $50, 50, 10, 4$ 
equidistant values of $s_{1}$ in $[5, 20]$ and $400, 19, 10, 1$ equidistant 
values of $\theta_{q}$ in $[0,1[$, respectively. 
For each choice the ordered eigenphases $\phi_{n}$ give the unfolded level 
spacings
\begin{align}
  \label{eq:spacings}
  s_{n} := \frac{N}{2\pi} (\phi_{n+1} - \phi_{n}).
\end{align}
Assuming an uncorrelated superposition of regular and chaotic subspectra 
corresponding to disjoint regular and chaotic regions in phase space, these 
spacings are expected to follow the Berry-Robnik distribution \cite{BeRo1984}.
The only relevant parameter of this distribution is the density of
regular states which semiclassically equals the relative
size of the regular region in phase space, $\Areg$; see Eq.~\eqref{eq:dors}.
This gives the standard Berry-Robnik distribution
\begin{align}
  \label{eq:berryRobnik}
  P_{\mathrm{BR}}(s) = \frac{\dd{}^{2}}{\dd{}s^{2}} \left \{ \exp(-\Areg s)
    \ \mathrm{erfc}\left( \frac{\sqrt{\pi}}{2} (1-\Areg) s \right ) \right \}.
\end{align}
For purely chaotic systems one has $\Areg=0$ such that the Wigner 
distribution $\Ps{c}(s) = (\pi s / 2) e^{-\pi s^{2} / 4}$ is recovered.
For  purely regular systems one has $\Areg=1$, giving the Poisson 
distribution $\Ps{r}(s) = \e^{- s}$.

\begin{figure}[t]
  \begin{center}
    \includegraphics[width=\linewidth]{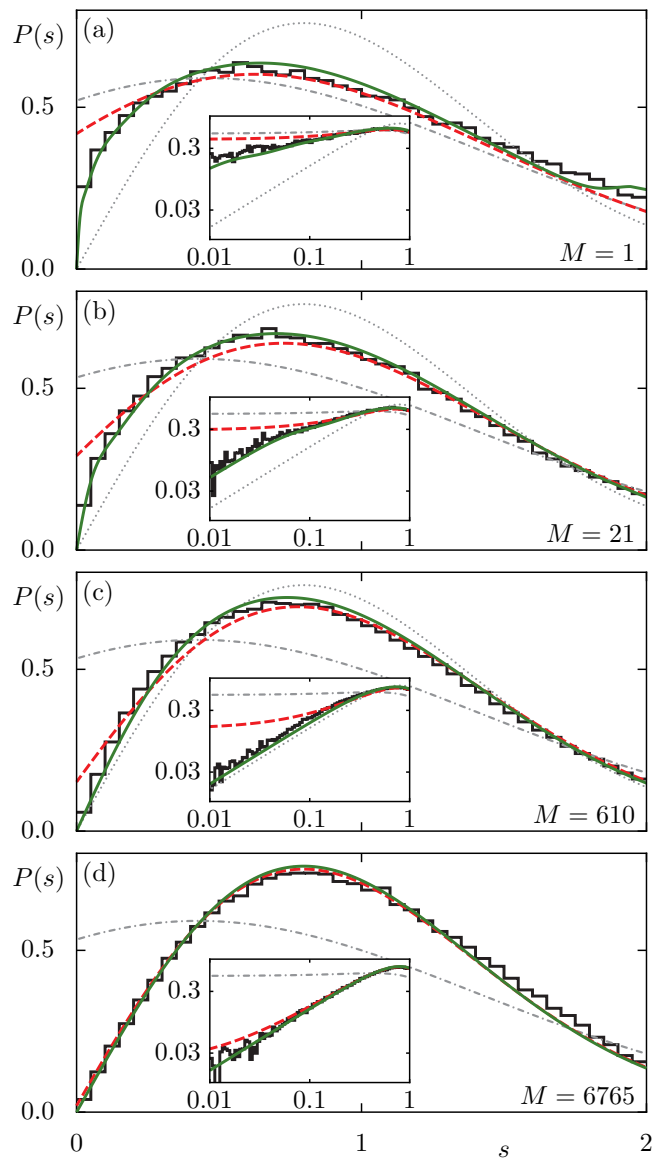}
    \caption{(Color online) Level-spacing distribution $P(s)$ of the model 
      system for $\heff \approx 1/13$. 
      The numerical data (black histograms) show a transition from the
      Berry-Robnik distribution (dash-dotted lines) to the Wigner distribution
      (dotted lines) with increasing system size $M=1$, $21$, $610$, and 
      $6765$ (a)-(d).
      These data are compared to the flooding-improved Berry-Robnik distribution
      (red dashed lines), Eq.~\eqref{eq:effBerryRobnik}, as well as to the
      flooding- and tunneling-improved Berry-Robnik
      distribution (green solid lines), Eq.~\eqref{eq:ftibr_dist}.
      The insets show the same distributions on a double logarithmic scale.
    }
    \label{fig:PS}
  \end{center}
\end{figure}

For the model systems introduced in Sec.~\ref{sec:modelSystem} 
one has $\Areg\approx 0.32$, such that 
Eq.~\eqref{eq:berryRobnik} predicts the same level-spacing distribution for 
all system sizes $M$. 
This is in contrast to our numerical findings, which show a transition of the 
level-spacing distribution from Berry-Robnik to Wigner statistics with 
increasing system size $M$ and fixed $\heff$.
In Fig.~\ref{fig:PS} numerical results for the level-spacing distribution of 
the model systems are shown as black histograms. 
For the case of only one unit cell [Fig.~\ref{fig:PS}(a)] the 
level-spacing distribution roughly follows the Berry-Robnik distribution 
(dash-dotted line).
With increase of the system size to $M=21$ unit cells [Fig.~\ref{fig:PS}(b)] 
the level-spacing 
distribution shows global deviations from the Berry-Robnik distribution.
For even larger system sizes [Figs.~\ref{fig:PS}(c) and \ref{fig:PS}(d)] 
we observe a transition to the Wigner 
distribution (dotted line).
This transition is caused by flooding of regular states, 
which we discuss in the following section.

\subsection{Flooding of regular states}
\label{sec:flooding}

We now show how the number of regular and chaotic states is
modified in the presence of flooding.
According to Eq.~\eqref{eq:bohr-sommerfeld} each regular state occupies an area
$\heff$ in phase space. Hence, the maximal number of quantizing tori $\mmax$ per
island is given by
\begin{align}
  \label{eq:mmax}
  \mmax = \left\lfloor \frac{\Areg}{\heff} + \frac{1}{2} \right\rfloor 
           \approx \frac{\Areg}{\heff}
\end{align}
and the quantum number $m$ runs from $0$ to $\mmax-1$.
Since we consider a chain of $M$ islands there are $M$ regular 
levels supported by the $m$th quantizing tori of the $M$ islands.
Hence, we semiclassically expect 
\begin{align}
  \label{eq:Nreg}
  \Nreg = \mmax M \approx \Areg M / \heff  
\end{align}
regular states supported by the $M$ regular islands of size $\Areg$.
The semiclassically expected density of regular states $\rhorsc$ is therefore 
given by the relative size of the regular region,
\begin{align}
  \label{eq:dors}
  \rhorsc := \frac{\Nreg}{N} \approx \Areg.
\end{align}
Similarly we expect $\Nch = N -\Nreg$ chaotic states and 
$\rhochsc = 1 - \rhorsc$.

Due to dynamical tunneling, regular and chaotic states are coupled. 
The average coupling of the regular states localizing on the $m$th quantizing 
tori to the chaotic states is given by the typical coupling $v_{m}$ 
\cite{BaKeLoMe2011}. 
It is determined by the tunneling rate $\gamma_{m}$ which describes the 
initial exponential decay of the $m$th regular WKB state 
to the chaotic sea,
\begin{align}
  \label{eq:effCoupling}
  v_{m} = \frac{N}{2\pi}\sqrt{\frac{\gamma_{m}}{\Nch}} = 
            \frac{1}{2\pi}\sqrt{\frac{\gamma_{m}}{\heff \rhochsc}}\sqrt{M}.
\end{align}
We compute the system specific tunneling rates $\gamma_{m}$ 
numerically \cite{BaKeLo2010}. 
They depend only on Planck's constant $\heff$ and the 
classical phase-space structure of one regular island, which are fixed in our 
investigations. 
Hence, the factor $\sqrt{\gamma_{m}/(\heff \rhochsc)}$ in 
Eq.~\eqref{eq:effCoupling} is constant for our model systems and the typical 
coupling $v_{m}$ is tunable by the system size $M$.
Note that the couplings $v$ used in Refs.~\cite{BaKeMo2005, BaKeMo2007} 
differ by the factor $\rhochsc$ from our definition, 
Eq.~\eqref{eq:effCoupling}, due to a different choice of dimensionless units.

In Ref.~\cite{BaKeMo2005} it was shown that in addition to the 
WKB quantization condition \eqref{eq:bohr-sommerfeld} regular states
exist on the $m$th quantizing tori only if the tunneling rate $\gamma_{m}$ is 
smaller than the inverse Heisenberg time 
of the chaotic subsystem, $\gamma_{m} < 1/\tauch$, 
Eq.~\eqref{eq:exCrit2}.
Using Eq.~\eqref{eq:effCoupling} and $\tauch=\heff/\dch=\Nch$ 
we rewrite this existence criterion in terms of the typical coupling,
\begin{align}
  \label{eq:exCrit3}
  v_{m} < \frac{1}{2\pi\rhochsc}.
\end{align}
If the existence criterion \eqref{eq:exCrit3} is fulfilled, the typical 
coupling of the WKB states on the $m$th quantizing tori is smaller than the 
chaotic mean level spacing and the corresponding regular eigenstates exist.
If $v_{m}$ increases beyond this threshold, the regular states on the 
$m$th quantizing tori effectively couple to an increasing number of spectrally 
close chaotic states.
Consequently the corresponding regular eigenstates disappear.
This process is called flooding of regular states \cite{BaKeMo2005, 
BaKeMo2007, LBphd, BaBiKe2012}. 
Thus for large typical couplings $v_{m}$ the number $\Ntil$ of regular states 
which actually exist in the regular islands is smaller than the 
semiclassically expected number $\Nreg$ of regular states. 
The quantizing tori of the $\Nreg - \Ntil$ regular states which violate 
Eq.~\eqref{eq:exCrit3} are flooded by chaotic states in phase space. 
Note that for our model systems the relation 
$v_{0} < v_{1} < v_{2} < \hdots$ holds, such that the quantizing 
tori are flooded in the order of decreasing quantum number $m$ from the border 
to the center of the regular islands.

\subsection{Flooding-improved Berry-Robnik distribution}
\label{sec:effBeRoStat}

We now introduce a flooding-improved Berry-Robnik distribution which takes the 
flooding of regular states into account. For that purpose we compute the 
density of regular states $\rhor$ in the presence of flooding. 
Starting from the semiclassically expected density of regular 
states, Eq.~\eqref{eq:dors}, and using Eqs.~\eqref{eq:heff} and 
\eqref{eq:Nreg} we obtain
\begin{align}
  \label{eq:rhorsc2}
  \rhorsc \approx \sum_{m=0}^{\mmax-1} \heff.
\end{align}
This expression shows that each quantizing torus semiclassically 
contributes one Planck cell to the density of states. 
To compute the density of regular states $\rhor$ in the presence of flooding 
we include only those quantizing tori in the sum in Eq.~\eqref{eq:rhorsc2} 
for which the existence criterion \eqref{eq:exCrit3} holds, and obtain 
\cite{BaKeMo2005, BaKeMo2007}
\begin{align}
  \label{eq:rhoApp}
  \rhor := \sum_{m=0}^{\mmax-1} \heff 
    \left[1 - \Theta\left(v_{m} - \frac{1}{2\pi\rhochsc}\right)\right].
\end{align}

\begin{figure}[b]
  \begin{center}
    \includegraphics[width=\linewidth]{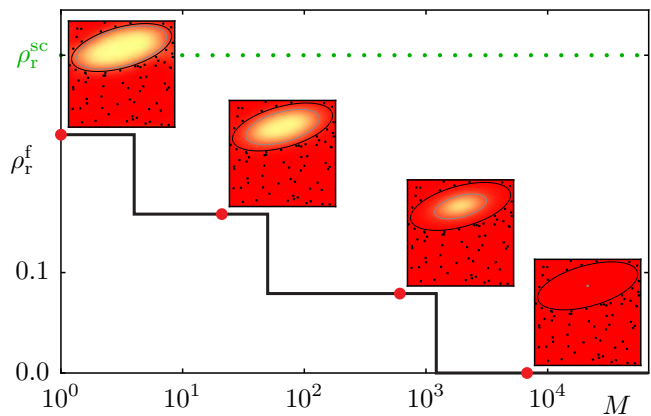}
    \caption{(Color online) Density of regular states $\rhor$ in the presence 
      of flooding, Eq.~\eqref{eq:rhoApp}, vs.\ the system size $M$ for 
      $\heff\approx 1/13$ on a logarithmic abscissa. 
      For comparison, the semiclassically expected density of regular states 
      $\rhorsc$ is shown (green dotted line).
      The insets illustrate the classical phase space where the gray tori 
      enclose the area $\rhor$ for $M=1$, $21$, $610$, and $6765$. 
      In addition, the averaged Husimi function of all chaotic eigenstates 
      folded into the first unit cell is shown.}
    \label{fig:AEff}
  \end{center}
\end{figure}

In Fig.~\ref{fig:AEff} the density of regular states $\rhor$ is shown for the 
example system at $\heff\approx 1/13$ versus the system size $M$ on a 
logarithmic abscissa. 
It decreases with increasing system size $M$ and has a step whenever a typical 
coupling $v_m$ equals $1/(2\pi\rhochsc)$. The semiclassical 
density of regular states $\rhorsc$ is an upper limit. 
In the spirit of Eq.~\eqref{eq:dors}, $\rhorsc \approx \Areg$, we interpret 
the density of regular states $\rhor$ in the presence of flooding by an area 
in phase space.
For $M=1$, $21$, $610$, and $6765$ the insets of Fig.~\ref{fig:AEff} show this 
area enclosed by a gray torus.
In addition the averaged Husimi functions 
(bright/yellow to darker/red color scale) illustrate 
that the surviving regular states are localized in this area $\rhor$, 
which decreases in the flooding limit $M\to\infty$. Already for the 
system with one unit cell, $M=1$, we find that 
$\rhor$ is smaller than its semiclassical expectation $\rhorsc$ because 
the outermost regular state of quantum number $m=3$ violates the 
existence criterion \eqref{eq:exCrit3}.

Note that the amount by which a regular state is flooded can also be described 
by smooth functions, e.g., by the fraction of regular states 
\cite{BaKeMo2007} or the asymptotic flooding weight \cite{BaBiKe2012}. 
However, they do not provide a significant advantage for our investigations.

To obtain a description of the level-spacing distribution which includes 
flooding, we now use prediction \eqref{eq:rhoApp} for the density of regular 
states $\rhor$ instead of the relative size of the regular 
region $\Areg$ as the relevant parameter in
Eq.~\eqref{eq:berryRobnik}. 
With $\rhoch := 1 - \rhor$ this leads to the flooding-improved Berry-Robnik 
distribution
\begin{align}
  \label{eq:effBerryRobnik}
  P_{\mathrm{fi}}(s) = \frac{\dd{}^{2}}{\dd{}s^{2}}
  \left \{ 
    \exp(-\rhor s) \ \mathrm{erfc}\left( 
      \frac{\sqrt{\pi}}{2} \rhoch s \right) 
  \right \},
\end{align}
which is our first main result. 
In Fig.~\ref{fig:PS} we compare the numerically determined nearest-neighbor 
level-spacing distribution to the analytical prediction 
(\ref{eq:effBerryRobnik}) for our model system. 
With increasing system size $M$ and fixed $\heff$ we find a global transition of the 
level-spacing distribution from Berry-Robnik to Wigner statistics.
This global transition is well described by the flooding-improved Berry-Robnik 
distribution, Eq.~(\ref{eq:effBerryRobnik}). 
It is a consequence of flooding, which reduces the density of regular states 
below its semiclassical expectation, $\rhor \le \rhorsc$.
According to Eq.~\eqref{eq:dors} this can be interpreted as a flooding 
induced decrease of the regular region in phase space.
In the limit $M \to \infty$ the regular islands are completely flooded and no 
regular state exists.
Hence, the Wigner distribution is obtained.

Note that even for the case of only one unit cell [see Fig.~\ref{fig:PS}(a)] 
non-zero couplings $v_{m}$ exist such that the numerical data are better 
described by the flooding-improved Berry-Robnik distribution, 
Eq.~\eqref{eq:effBerryRobnik}, than by Eq.~\eqref{eq:berryRobnik}. 

At small spacings deviations between numerical data and the flooding-improved 
Berry-Robnik distribution are visible.
They occur due to level repulsion between the surviving regular and the 
chaotic levels, which is not considered within this approach and will be 
incorporated in the following section.

\subsection{Flooding- and tunneling-improved Berry-Robnik distribution}
\label{sec:complTheory}

We now unify the flooding-improved Berry-Robnik distribution,
Eq.~\eqref{eq:effBerryRobnik}, with the tunneling-improved Berry-Robnik 
distribution \cite{BaKeLoMe2011}.
The resulting flooding- and tunneling-improved Berry-Robnik distribution
allows us to describe both the effect of flooding and the 
power-law level repulsion at small spacings.
The derivation is done along the lines of Ref.~\cite{BaKeLoMe2011}.
We incorporate the effects of 
flooding into this theory by replacing the number of 
regular states $\Nreg$ with the number of surviving regular states 
$\Ntil$ which fulfill the existence criterion \eqref{eq:exCrit3}. 
The other regular states, which fulfill the WKB quantization condition 
\eqref{eq:bohr-sommerfeld} but have strong couplings to the chaotic sea, 
$v_{m} > 1/(2\pi\rhochsc)$, are assigned to the chaotic subspectrum.
Level repulsion is then modeled by accounting for 
the small tunneling couplings $v_m$ between the $\Ntil$ surviving regular 
states and the chaotic states perturbatively.

Following Refs.~\cite{BeRo1984, PoNa2007, BaRo2010, BaKeLoMe2011}
the flooding- and tunneling-improved 
Berry-Robnik distribution $P_{\mathrm{fti}}(s)$ consists of three distinct 
parts:
\begin{align}
  \label{eq:partdist}
  P_{\mathrm{fti}}(s) = \ps{s}{r-r} + \ps{s}{c-c} + \ps{s}{r-c}.
\end{align}
Here $\ps{s}{r-r}$ describes the contribution of level spacings between two 
regular levels, $\ps{s}{c-c}$ the contribution of level spacings between 
two chaotic levels, and $\ps{s}{r-c}$ the contribution of level spacings 
formed by one regular and one chaotic level in the superposed spectrum. 
In our model systems the number of quantizing tori $\mmax$ is small, e.g., 
$\mmax \approx 4$, and the $M$ regular levels with the same quantum number 
$m$ are equispaced with distance $N/M$ in the unfolded spectrum 
\cite{BrHo1991}. 
Hence, the regular levels do not follow the generic Poissonian behavior 
occurring for large $\mmax$, but are well separated,
\begin{align}
  \label{eq:regSpace}
  \ps{s}{r-r}\approx0.
\end{align}
Furthermore, 
\begin{align}
  \label{eq:chSpace}
  \ps{s}{c-c} = \Ps{c}(s) [1-\rhorsc s],
\end{align}
where $\Ps{c}(s)$ is the Wigner distribution, which describes the probability
of finding a spacing $s$ in the chaotic subspectrum. The second factor 
$[1-\rhorsc s]$ describes the probability of finding a gap in the regular 
subspectrum. For the third term in Eq.~\eqref{eq:partdist} one finds 
\cite{BaKeLoMe2011}
\begin{align}
  \label{eq:rcdist}
  \ps{s}{r-c} = p^{(0)}_{\mathrm{r-c}}(s) \frac{1}{\Ntil} 
  \sum_{m=0}^{\Ntil-1} \frac{\tilde{v}_{m}}{v_{m}} 
  X\left( \frac{s}{2v_{m}} \right),
\end{align}
with $X(x):=\sqrt{\pi/2}\, x\exp(-x^{2}/4)I_{0}(x^{2}/4)$, where $I_{0}$ 
is the zeroth-order modified Bessel function and 
$\tilde{v}_{m}=v_{m}/\sqrt{1-2\pi(\rhochsc v_{m})^2}$.
The contribution of the zeroth-order regular-chaotic spacings, 
$p^{(0)}_{\mathrm{r-c}}(s)$, is given by 
\begin{align}
  \label{eq:zeroth-r-c}
  p^{(0)}_{\mathrm{r-c}}(s) = 2 \rhochsc \rhorsc 
  \exp\left( -\frac{\pi (\rhochsc s)^{2}}{4}\right).
\end{align}
Using Eqs.~\eqref{eq:regSpace}, \eqref{eq:chSpace}, and \eqref{eq:rcdist} 
in Eq.~\eqref{eq:partdist}, we obtain the flooding and tunneling improved 
Berry-Robnik distribution
\begin{align}
  \label{eq:ftibr_dist}
  P_{\mathrm{fti}}(s) = \Ps{c}(s) [1-\rhorsc s] + p^{(0)}_{\mathrm{r-c}}(s) 
  \frac{1}{\Ntil} \sum_{m=0}^{\Ntil-1} \frac{\tilde{v}_{m}}{v_{m}} 
  X\left( \frac{s}{2v_{m}} \right),
\end{align}
which is our final result.
In Eq.~\eqref{eq:ftibr_dist} one has to sum over the $\Ntil \le \Nreg$ regular 
states, which fulfill the existence criterion \eqref{eq:exCrit3}. 
This selection of the regular states takes flooding into account. 
In addition power-law level repulsion at small spacings is described 
by the last term of Eq.~\eqref{eq:ftibr_dist}.

In Fig.~\ref{fig:PS} we compare the numerical results for the 
level-spacing distribution to the flooding- and tunneling-improved 
Berry-Robnik distribution, Eq.~\eqref{eq:ftibr_dist} (green solid lines). 
We find excellent agreement.
The global transition of the level-spacing distribution from the Berry-Robnik 
distribution in Fig.~\ref{fig:PS}(a) for a system with one unit cell to 
the Wigner distribution in Fig.~\ref{fig:PS}(d) for a system with $M=6765$
is well described. 
This transition is caused by the disappearance of regular states due to 
flooding.
Furthermore, the flooding- and tunneling-improved Berry-Robnik distribution, 
Eq.~\eqref{eq:ftibr_dist}, reproduces the power-law level repulsion 
of $P(s)$ at small spacings, which is caused by small tunneling splittings 
between the surviving regular and chaotic states. 
This can be seen particularly well in the double logarithmic insets of 
Fig.~\ref{fig:PS}.

\subsection{Limiting cases}
\label{sec:compare}

Depending on the interplay between the flooding limit $M\to\infty$ and the 
semiclassical limit $\heff\to 0$, we identify three cases 
in which the tunneling corrections of Sec.~\ref{sec:complTheory} 
are insignificant. 

Case (i) is the flooding limit with fixed $\heff$ and $M\to \infty$ in 
which all regular states are flooded. 
Asymptotically one obtains the Wigner distribution; see Fig.~\ref{fig:PS}(d). 
Note that a further increase of the system size after all regular states have 
been flooded completely, may lead to the localization of chaotic eigenstates 
which affects spectral statistics \cite{Iz1990, ChSh1995}.

Case (ii) considers the semiclassical limit $\heff \to 0$ for fixed system 
sizes $M$. 
In this case both flooding and tunneling corrections vanish due to 
exponentially decreasing tunneling couplings. 
Hence, the spacing statistics tend toward
the standard Berry-Robnik distribution, Eq.~\eqref{eq:berryRobnik}
\cite{BaKeLoMe2011, PrRo1994}. 

In case (iii) the semiclassical limit $\heff \to 0$ and the 
flooding limit $M \to \infty$ are coupled such that $\rhor$ 
is constant and smaller than $\rhorsc$.
In this limit flooding is present yet the tunneling corrections at small 
spacings vanish.
In this case the spacing statistics are given by the flooding-improved 
Berry-Robnik distribution, Eq.~\eqref{eq:effBerryRobnik}.

\begin{figure}[!t]
  \begin{center}
    \includegraphics[width=\linewidth]{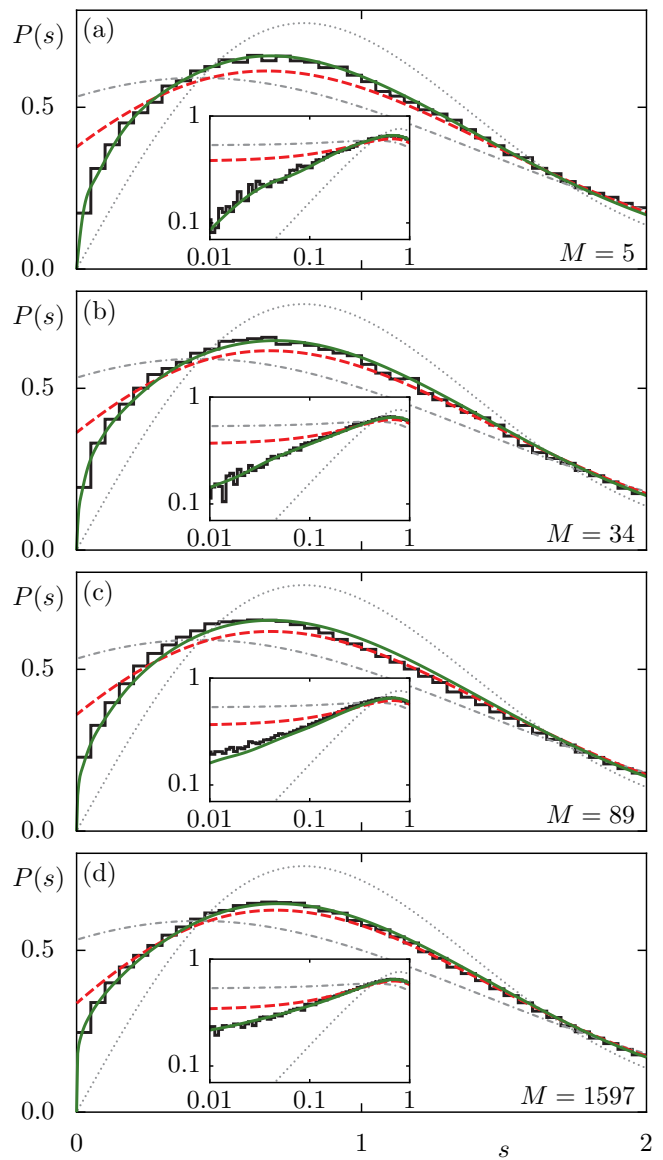}
    \caption{(Color online) Level-spacing distribution $P(s)$ of the model 
      system at fixed density of regular states $\rhor \approx 0.58 \rhorsc$. 
      The numerical data (black histograms) are compared to the 
      flooding-improved Berry-Robnik distribution (red dashed lines), 
      Eq.~\eqref{eq:effBerryRobnik}, as well as to the flooding- and 
      tunneling-improved Berry-Robnik distribution (green solid lines), 
      Eq.~\eqref{eq:ftibr_dist}, for $(M, N) = (5, 63)$, $(34, 735)$, 
      $(89, 2458)$, and $(1597, 77643)$ (a)-(d), corresponding to 
      $\heff\approx1/13$, $1/22$, $1/28$, and $1/49$, respectively.
      For comparison, the Wigner distribution (dotted lines) and the 
      Berry-Robnik distribution (dash-dotted lines) are shown.
      The insets show the same distributions on a double logarithmic scale.
      }
    \label{fig:ps_fconst}
  \end{center}
\end{figure}

In Fig.~\ref{fig:ps_fconst} we illustrate spectral statistics in the limit of 
case (iii).
We consider the model systems for $(M, N) = (5, 63)$, $(34, 735)$, 
$(89, 2458)$, $(1597, 77643)$ such that the density of regular states is 
fixed, $\rhor\approx 0.58 \rhorsc$, and $\heff \approx 1/13, 1/22, 1/28, 1/49$
decreases.
Both the numerical data and the flooding- and tunneling-improved 
Berry-Robnik distribution, Eq.~\eqref{eq:ftibr_dist}, tend toward the 
flooding-improved Berry-Robnik distribution, Eq.~\eqref{eq:effBerryRobnik}.
The vanishing influence of the tunneling corrections at small spacings is 
particularly visible in the insets, which show the spacing distributions on a 
double logarithmic scale.

\section{Summary}
\label{sec:outlook}

In this paper we study the impact of flooding on the level-spacing 
distribution $P(s)$ for systems with a mixed phase space. 
Numerically we find a transition from Berry-Robnik to Wigner statistics with 
increasing density of states and fixed $\heff$. 
We explain this transition by the flooding of regular islands. It reduces the 
density of regular states $\rhor$ below its semiclassical expectation 
$\rhorsc$ which can be interpreted as a flooding induced decrease of the 
regular region in phase space.
Taking this into account we derive a flooding-improved Berry-Robnik 
distribution, which reproduces the observed transition of the level-spacing 
statistics. 
We unify this prediction with the tunneling-improved Berry-Robnik distribution 
\cite{BaKeLoMe2011} which includes power-law level repulsion. This gives the
flooding- and tunneling-improved Berry-Robnik distribution, which shows 
excellent agreement with numerical data. 

In order to demonstrate the effect of flooding on spectral statistics, we
investigated model systems with a simple phase-space structure. However, we
expect that flooding has similar consequences for systems with a generic
phase space, which may contain several regular and chaotic regions as well as
nonlinear resonances, larger hierarchical regions, and partial transport
barriers. This expectation is based on the fact that even for generic systems
tunneling couplings between classically disjoint regions exist. Hence,
increasing the density of states for fixed $\heff$ should still lead to a
transition of the level-spacing distribution from Berry-Robnik to Wigner
statistics.

\begin{acknowledgments}
We thank Roland Ketzmerick and Lars Bittrich for stimulating discussions.
Furthermore, we acknowledge support by the Deutsche Forschungsgemeinschaft 
within the Forschergruppe 760 ``Scattering Systems with Complex Dynamics.''
\end{acknowledgments}

\begin{appendix}
\section*{Appendix: Periodic band matrix}
\label{sec:perBandMat}

Our aim is to calculate the eigenphases $\phi_{n}$ from 
Eq.~\eqref{eq:eigvalProb} numerically up to Hilbert-space dimension 
$N\approx10^{5}$. 
This is possible due to a band-matrix algorithm \cite{BaKe2000} which was 
used in Refs.~\cite{BaKeMo2005, BaKeMo2007} and is presented in the following.

We start with the matrix representation of the symmetrized 
time-evolution operator, 
\begin{align}
\op{U}^{\hf} = \e^{-\frac{\im}{2\heffbar}T(\hat{p})}\,
\e^{-\frac{\im}{\heffbar}V(\hat{q})}\,\e^{-\frac{\im}{2\heffbar}T(\hat{p})},
\end{align}
in the basis of the discretized position states $\ket{q_{k}}$,
\begin{align}
  \label{eq:uhfPos}
  U^{\hf}_{k,l} := \bra{q_{k}}\op{U}^{\hf}\ket{q_{l}},
\end{align}
with $q_{k} = \heff (k+\theta_{p}-\tfrac{1}{2})$ and $k,l=0, 1, \dots, N-1$.
This matrix has dominant contributions around the diagonal and in the 
upper right and lower left corners, i.e., $\op{U}^{\hf}$ can be approximated 
by a periodic band matrix.
For our model systems the width of the band depends on the extrema 
of $V'(q)$.
The essential step for computing the eigenvalues of $U^{\hf}$ is to 
find a similarity transformation from this periodic band matrix to a band 
matrix. For the Hermitian case a similar idea was used in 
Ref.~\cite{Kra1996}.

Since the kicking potential is symmetric 
about $q=0$ for 
$M(\theta_p+N/2)\in\mathbb{Z}$, $V(q_{l}) = V(q_{N-l})$, we find
\begin{align}
  \label{eq:uhfSym}
  U^{\hf}_{k,l} = U^{\hf}_{N-l, N-k}.
\end{align}
Hence, the set of eigenvectors $\ket{\phi_{n}}$, for which we choose the phase 
such that $\scal{q_{0}}{\phi_{n}} = \scal{q_{0}}{\phi_{n}}^{*}$, satisfies
\begin{align}
  \label{eq:phiSym}
  \scal{q_{l}}{\phi_{n}} &= \scal{q_{N-l}}{\phi_{n}}^{*},
\end{align}
where the star denotes the complex conjugation and $l$ runs from $1$ to $N-1$. 
Based on these relations it is possible to find a 
unitary transformation $\op{A}$ to a set of purely real vectors 
$\ket{\psi_{n}} := \op{A} \ket{\phi_{n}}$, given by
\begin{align}
  \label{eq:unitTrafo}
  \scal{q_{0}}{\psi_{n}} &= \scal{q_{0}}{\phi_{n}},\\
  \scal{q_{2k-1}}{\psi_{n}} &= \frac{1}{\sqrt{2}}(\scal{q_{k}}{\phi_{n}} + 
  \scal{q_{N-k}}{\phi_{n}}),\\
  \scal{q_{2k}}{\psi_{n}} &= \frac{1}{\im\sqrt{2}}(\scal{q_{k}}{\phi_{n}} - 
  \scal{q_{N-k}}{\phi_{n}}),\\
  \scal{q_{N-1}}{\psi_{n}} &= \scal{q_{N/2}}{\phi_{n}}.
\end{align}
Here, $k$ runs from $1$ to $(N-1)/2$ for odd $N$ 
or from $1$ to $(N-2)/2$ for even $N$ and the last row has to be considered
only for even $N$.

We now define a new operator $\op{W}$, given by the unitary 
transformation of $\op{U}^{\hf}$ with $\op{A}$,
\begin{align}
  \label{eq:wDef}
  \op{W} := \op{A}\ \op{U}^{\hf} \op{A}^{-1}.
\end{align}
Its matrix representation $\mat{W}$ in the basis of position states 
$\ket{q_{k}}$ has a banded structure with twice the bandwidth of the matrix 
$\mat{U}^{\hf}$ but without components in the upper right and lower left 
corners.
Furthermore it is symmetric,
\begin{align}
  \label{eq:wSym}
  W_{k,l} = W_{l,k},
\end{align}
with complex matrix elements $W_{k,l}$.
The unitary transformation \eqref{eq:wDef} leads to a new eigenvalue problem 
with the same eigenphases as in Eq.~\eqref{eq:eigvalProb},
\begin{align}
  \label{eq:newEigvalProb}
  \op{W} \ket{\psi_{n}} = \e^{\im \phi_{n}} \ket{\psi_{n}}.
\end{align}
Numerical standard libraries provide methods for the eigenvalue computation 
of only real symmetric or complex Hermitian band matrices
but not for unitary band matrices such as $\mat{W}$.
Hence, we first calculate the real part of the eigenvalues, following from 
$\Re\{\mat{W}\} \ket{\psi_{n}} = \cos\phi_{n}\ket{\psi_{n}}$, and afterwards 
the imaginary part from 
$\Im\{\mat{W}\} \ket{\psi_{m}} = \sin\phi_{m}\ket{\psi_{m}}$. This is possible 
since the eigenvectors $\ket{\psi_{n}}$ are purely real. 
From these results one can recover the eigenphases $\phi_n$ by the 
requirement $\cos^2\phi_n + \sin^2\phi_m = 1$. The corresponding eigenfunctions 
$\ket{\psi_n}$ can be obtained from Eq.~\eqref{eq:newEigvalProb} by the method 
of inverse iteration using an LU decomposition. 
By the mapping of the original eigenvalue problem Eq.~\eqref{eq:eigvalProb}
to the band matrix form Eq.~\eqref{eq:newEigvalProb}, 
it is possible to compute both eigenvalues and
eigenfunctions with a numerical effort scaling as $N^2$ in contrast to the 
standard diagonalization procedures which scale as $N^3$.

\end{appendix}

\end{document}